\documentclass[prb,aps,twocolumn]{revtex4}
\usepackage{graphicx}
\usepackage{dcolumn}
\usepackage{bm}
\usepackage{color}
\usepackage{ulem}

\begin{document}
\newcommand{\bR}{\mbox{\boldmath $R$}}
\newcommand{\tr}[1]{\textcolor{black}{#1}}
\newcommand{\trs}[1]{\textcolor{black}{\sout{#1}}}
\newcommand{\tb}[1]{#1}
\newcommand{\tbs}[1]{\textcolor{black}{\sout{#1}}}
\newcommand{\Ha}{\mathcal{H}}
\newcommand{\mh}{\mathsf{h}}
\newcommand{\mA}{\mathsf{A}}
\newcommand{\mB}{\mathsf{B}}
\newcommand{\mC}{\mathsf{C}}
\newcommand{\mS}{\mathsf{S}}
\newcommand{\mU}{\mathsf{U}}
\newcommand{\mX}{\mathsf{X}}
\newcommand{\sP}{\mathcal{P}}
\newcommand{\sL}{\mathcal{L}}
\newcommand{\sO}{\mathcal{O}}
\newcommand{\la}{\langle}
\newcommand{\ra}{\rangle}
\newcommand{\ga}{\alpha}
\newcommand{\gb}{\beta}
\newcommand{\gc}{\gamma}
\newcommand{\gs}{\sigma}
\newcommand{\vk}{{\bm{k}}}
\newcommand{\vq}{{\bm{q}}}
\newcommand{\vR}{{\bm{R}}}
\newcommand{\vQ}{{\bm{Q}}}
\newcommand{\vga}{{\bm{\alpha}}}
\newcommand{\vgc}{{\bm{\gamma}}}
\newcommand{\Ns}{N_{\text{s}}}
\newcommand{\vx}{{\bf r}}
\newcommand{\vG}{{\bf G}}
\newcommand{\avrg}[1]{\left\langle #1 \right\rangle}
\newcommand{\eqsa}[1]{\begin{eqnarray} #1 \end{eqnarray}}
\newcommand{\eqwd}[1]{\begin{widetext}\begin{eqnarray} #1 \end{eqnarray}\end{widetext}}
\newcommand{\hatd}[2]{\hat{ #1 }^{\dagger}_{ #2 }}
\newcommand{\hatn}[2]{\hat{ #1 }^{\ }_{ #2 }}
\newcommand{\wdtd}[2]{\widetilde{ #1 }^{\dagger}_{ #2 }}
\newcommand{\wdtn}[2]{\widetilde{ #1 }^{\ }_{ #2 }}
\newcommand{\cond}[1]{\overline{ #1 }_{0}}
\newcommand{\conp}[2]{\overline{ #1 }_{0#2}}
\newcommand{\nn}{\nonumber\\}
\newcommand{\cdt}{$\cdot$}
\newcommand{\bra}[1]{\langle#1|}
\newcommand{\ket}[1]{|#1\rangle}
\newcommand{\braket}[2]{\langle #1 | #2 \rangle}
\newcommand{\bvec}[1]{\mbox{\boldmath$#1$}}
\newcommand{\blue}[1]{{#1}}
\newcommand{\bl}[1]{{#1}}
\newcommand{\bn}[1]{\textcolor{black}{#1}}
\newcommand{\rr}[1]{{#1}}
\newcommand{\rt}[1]{\textcolor{red}{#1}}
\newcommand{\mgt}[1]{\textcolor{black}{#1}}
\newcommand{\mg}[1]{#1}
\newcommand{\red}[1]{{#1}}
\newcommand{\fj}[1]{{#1}}
\newcommand{\green}[1]{{#1}}
\newcommand{\gr}[1]{\textcolor{black}{#1}}
\definecolor{green}{rgb}{0,0.5,0.1}
\definecolor{blue}{rgb}{0,0,0.8}
\preprint{APS/123-QED}

\title{{\it Ab Initio} Downfolding Study of the Iron-based Ladder Superconductor BaFe$_2$S$_3$
}
\author{Ryotaro Arita$^{1,2}$, Hiroaki Ikeda$^3$, Shiro Sakai$^1$ and Michi-To Suzuki$^1$
}
\affiliation{$^1$RIKEN Center for Emergent Matter Science, 2-1 Hirosawa, Wako, Saitama 351-0198, Japan}
\affiliation{$^2$ERATO Isobe Degenerate $\pi$-Integration Project, Tohoku University, Aoba-ku, Sendai 980-8578,
Japan}
\affiliation{$^3$Department of Physics, Ritsumeikan University, 1-1-1 Noji-higashi, Kusatsu, Shiga 525-8577, Japan}
\date{\today}

\begin{abstract}
Motivated by the recent discovery of superconductivity in the iron-based ladder compound BaFe$_2$S$_3$ under high pressure,
we derive low-energy effective Hamiltonians from first principles. 
We show that the complex band structure around the Fermi level is represented only by the Fe 3$d_{xz}$ (mixed with 3$d_{xy}$) and 3$d_{x^2-y^2}$ 
orbitals. The characteristic band degeneracy allows us to construct a four-band model with the band unfolding approach. 
We also estimate the interaction parameters 
and show that the system is more correlated than the 1111 family of iron-based superconductors. 
Provided the superconductivity is mediated by spin fluctuations, the $3d_{xz}$-like band plays
an essential role, and the gap function changes its sign between the Fermi surface 
around the $\Gamma$ point and that 
around the Brillouin-zone boundary.
\end{abstract}

\maketitle
\section{Introduction}
Since the discovery of superconductivity in fluorine doped LaFeAsO\cite{Hosono},  a variety of iron pnictides and
chalcogenides have been found to exhibit superconductivity with high transition temperatures ($T_c$). 
In these compounds, Fe ions commonly form a two-dimensional (2D) network. 
This fact raises an intriguing and fundamental question whether the square
network is essential for the high $T_c$ superconductivity and what happens in different geometries.
In the case of the cuprates, it was discovered that the ladder compound (Sr,Ca)$_{14}$Cu$_{24}$O$_{41}$
becomes a superconductor under pressure $\sim$ 3 GPa.\cite{Uehara}
This experimental observation has stimulated various theoretical studies on 
superconductivity in quasi-1D systems. 
Interestingly, it has been recently found that a ladder compound BaFe$_2$S$_3$ becomes a 
superconductor under high pressure $\sim$ 10GPa.\cite{Ohgushi}  
In the phase diagram, the superconducting phase resides next to a magnetic insulating phase, 
in which the spin correlation is antiferromagnetic along the ladder and
ferromagnetic along the rung. The maximum $T_c$ is as high as 14 K.
The superconducting transition in 
BaFe$_2$S$_3$ is of great interest, since
we may have a chance to pin down the origin/mechanism of the high $T_c$ superconductivity in 
iron-based superconductors (FeSC) by
investigating the commonalities and differences between the 2D and quasi-1D systems.

Recently, motivated by the experimental works on superconductivity in the single-layer 
potassium-doped iron selenide (110) film (which can be viewed as a weakly coupled ladder system)\cite{FeSeladder}
or ladder compounds such as BaFe$_2$Se$_3$\cite{Caron,Krzton,Lei,Saparov,Caron2,Nambu} and CsFe$_2$Se$_3$\cite{Du}
(which do not exhibit superconductivity), a variety of theoretical studies have been 
reported.\cite{WeiLi,Nekrasov,Dagotto1, Dagotto2, DagottoPRB1, DagottoPRB2,DagottoRMP}
However, studies for BaFe$_2$S$_3$ based on {\it ab initio} calculation are yet to be performed.
In this study, we derive low-energy effective Hamiltonians for BaFe$_2$S$_3$ from first principles, focusing 
on the Fe 3$d$ bands around the Fermi level ($E_F$).
We find that a $d_{xz}$-like orbital, a linear combination of  $d_{xz}$ and $d_{xy}$, forms two 
Fermi (electron) pockets around $k_z$=$0$ and a
$d_{x^2-y^2}$-like orbital forms a Fermi (hole) pocket around $k_z$=$\pi$.
The effective energy bands obtained with the two orbital models are further depicted in the unfolded Brillouin zone (BZ), in which one of the electron pockets is placed around the BZ boundary of $k'_z$=$\pi$.
Since the magnetic instability is strong at $q'_z \sim \pi$ in the extended BZ, 
the $d_{xz}$-like orbital should be active for 
spin-fluctuation-mediated superconductivity. We also estimate the values of interaction parameters in the
effective Hamiltonian, such as the Hubbard $U$ and Hund's coupling $J$. 
We show that BaFe$_2$S$_3$ is more strongly correlated than the 1111 compounds.

\section{Crystal structure}
\begin{figure}
\begin{center}
\includegraphics[width=6.5cm]{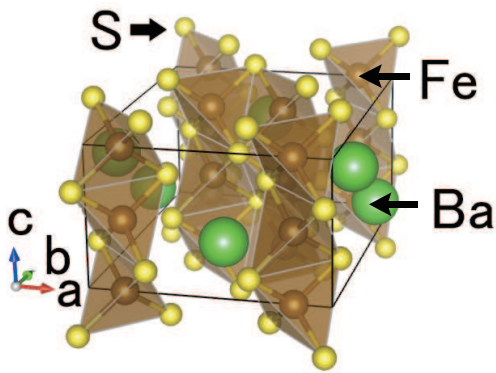}
\caption{(color online):
Crystal structure of BaFe$_2$S$_3$.
}
\label{crystal}
\end{center}
\end{figure}

In Fig. \ref{crystal}, we show the crystal structure of BaFe$_2$S$_3$. We see that Fe atoms
form two-leg ladders running along the $c$-axis. In the following calculation
for ambient pressure, we used 
the lattice constants $a$, $b$, $c$ and the atomic positions of Ba, Fe and S
reported in Ref.~\onlinecite{Ohgushi2}.
Namely, the lattice constant $a$, $b$, and $c$ are 8.78, 11.23 and 5.29 \AA,
respectively. The space group is Cmcm, and the atomic positions of
Ba(4c), Fe(8e), S(4c) and S(8g) are (0.0, 0.686, 0.25), (0.154, 0.0, 0.0),
(0.0, 0.116, 0.25) and (0.208, 0.378, 0.25), respectively.
In the phase diagram of temperature ($T$) and pressure ($P$), the superconducting phase 
has a dome-like shape and $T_c$ is highest around $P$=12.4 GPa.
For $P$=12.4 GPa, $a$, $b$ and $c$ shrink to 96.0\%, 92.0\% and 96.6\% of those 
at ambient pressure, respectively\cite{Ohgushi2,SuzukiPRB}.  Since the atomic configuration under pressure is yet to be
reported, in the present calculation, we just change the lattice constants.

\section{Band structure, Fermi surface and projected density of states}
We performed density functional calculation for the experimental crystal structure\cite{notestructure} with the
{\sc quantum espresso} package.\cite{Espresso}
Here we employ the exchange correlation functional
by Perdew {\it et al.}\cite{PBE96}. The wave functions are expanded
by plane waves up to a cutoff energy of 40 Ry, and an 8$\times$8$\times$8 $k$-mesh
in the first Brillouin zone is used.

Based on the band calculation, we further constructed Wannier functions for these bands
using the {\sc wannier90} package\cite{wannier90}. 
The resulting spreads of the $d_{3z^2-r^2}, d_{xz}, d_{yz}, d_{x^2-y^2}$, and $d_{xy}$ orbitals are
2.35, 3.51, 3.13, 2.40 and 3.14 \AA$^2$ for ambient pressure, and
2.91, 4.01, 3.35, 2.73 and 3.43 \AA$^2$ for $P=12.4$ GPa.
If we compare the Wannier spreads of the 2D FeSC
(Table III in Ref.~\onlinecite{Miyake}), we see that 
at $P$=0 the Wannier functions are as localized as those of LiFeAs, but at $P$=12.4 GPa they are more delocalized as those of BaFe$_2$As$_2$.

\begin{figure}
\centering
\includegraphics[width=8cm]{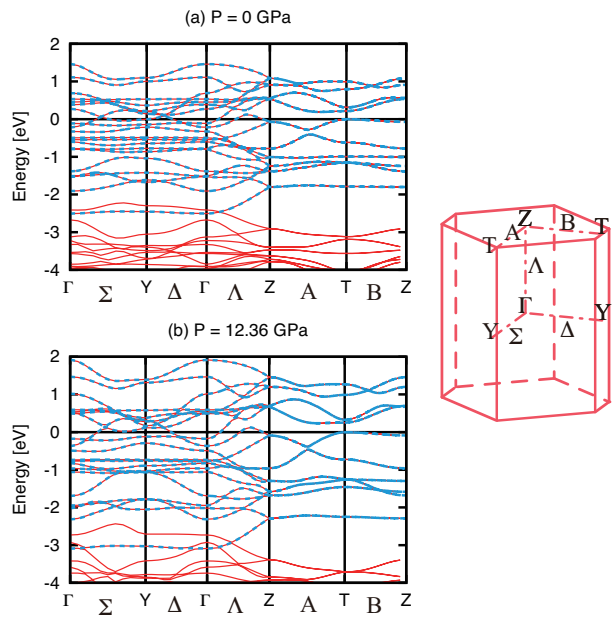}
\caption{(color online):
Band structure for ambient pressure (a) and
$P$=12.4 GPa (b). The original (Wannier interpolated)
band dispersion is shown by red solid (blue dashed) curves.
In the inset, we show the Brillouin zone.
}
\label{band}
\end{figure}

In Fig. \ref{band}, we show the band dispersion obtained by the
Wannier interpolation (blue dashed curves)
with that from the density functional calculation (red solid curves). Since the unit cell contains four Fe atoms, there are twenty Fe 3$d$ bands around the Fermi level.
We see that the band width of the Fe 3$d$ states for $P$=12.4 GPa is larger than that for 
$P$=0 by $\sim 25\%$. 
We list the resulting hopping integrals
in the Supplemental Material.\cite{suppl}
As is expected from the crystal structure shown in Fig. \ref{crystal}, the band structure is dispersive along the $k_z$ axis (from 
the $\Gamma$ to $Z$ point). The transfer integral is largest between the nearest neighbor
$d_{3z^2-r^2}$ orbitals along the ladder (the $c$-axis), and its amplitude is
$\sim$ 0.56 eV. For $P$=12.4 GPa, it becomes 0.64 eV. These values are larger than
the nearest-neighbor transfer integrals for the 2D FeSC (see Tables IV-VII in Ref.~\onlinecite{Miyake}).
 On the other hand, the inter-ladder transfer integrals are small, so that the band structure of BaFe$_2$S$_3$ is quasi-one dimensional along the $k_z$ axis. Nevertheless, since the dominant transfer integrals along the $c$ axis are large, the band width of BaFe$_2$S$_3$ turns out to be a similar value to that of 2D FeSC (see Fig. 4 in Ref.~\onlinecite{Miyake}).
As for the onsite energies of $d_{3z^2-r^2}, d_{xz}, d_{yz}, d_{x^2-y^2}$, and $d_{xy}$, they 
are 7.44, 8.00, 8.03, 7.69, and 7.95 eV (where $E_F$=8.19 eV) for $P$=0, 
and 9.37, 10.02, 9.98, 9.62 and 9.89 eV (where $E_F$= 10.22 eV) for $P$=12.4 GPa. 
The size of the crystal field splitting is similar to that of the 2D FeSC.

\begin{figure}
\centering
\includegraphics[width=7cm]{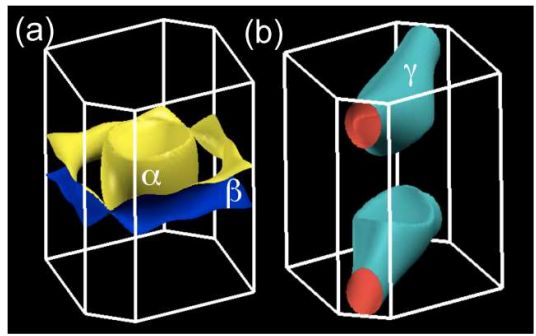}
\caption{(color online):
Fermi pockets $\alpha$ and $\beta$ (a) and $\gamma$ (b) of BaFe$_2$S$_3$ for $P$=0.
}
\label{fermi}
\end{figure}

In Fig.\ref{fermi}, we show the Fermi surface for $P$=0.
There are two electron pockets around $k_z=0$, hereafter called $\alpha$ and $\beta$, and one
hole pocket around $k_z=\pi$, called $\gamma$. 
For $P$=12.4 GPa, the $\beta$ pocket significantly shrinks, 
while the $\alpha$ pocket is quite robust against the external pressure.
The energy band lying just below $E_F$(=0 eV hereafter) around the $T$ points at $P$=0 crosses the Fermi level under pressure, resulting in another tiny pocket for $P$=12.4 GPa.

\begin{figure}
\centering
\includegraphics[width=8cm]{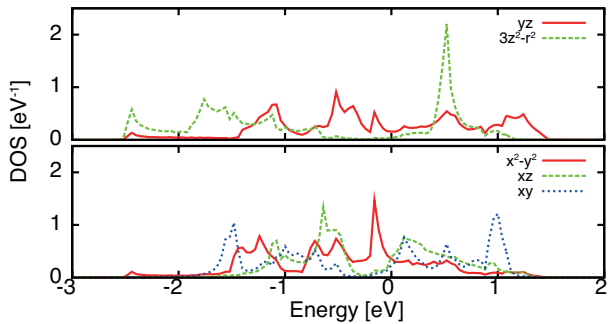}
\caption{(color online):
Projected density of states for $P$=0. While the amplitude of $d_{yz}$ and $d_{3z^2-r^2}$ 
around $E_F$ is not large, the contributions from $d_{x^2-y^2}$, $d_{xz}$ and $d_{xy}$ are
significant.
}
\label{pdos}
\end{figure}

We plot the projected density of states for the $3d$ orbitals 
in Fig.~\ref{pdos}. 
We see that $d_{3z^2-r^2}$ does not contribute to the low-energy states. On the other hand, the Fermi pockets $\alpha$, $\beta$, and $\gamma$ are mainly composed of $d_{xy}$, $d_{xz}$ and $d_{x^2-y^2}$, respectively. The contribution of $d_{yz}$ to these pockets is subdominant.
As for the electron filling $n$ of $d_{3z^2-r^2}, d_{xz}, d_{yz}, d_{x^2-y^2}$, and $d_{xy}$,
they are 0.61, 0.54, 0.56, 0.76, and 0.50 for $P$=0, and 
0.60, 0.51, 0.62, 0.76, and 0.53 for $P$=12.4 GPa.
Since $n$ of the $d_{x^2-y^2}$ orbital is quite high, $d_{x^2-y^2}$ should be
inactive magnetically.   
It is interesting to note that $n$ of $d_{3z^2-r^2}$ in the 2D FeSC
is $\sim$ 0.75 (see Table II in Ref.~\onlinecite{Miyake}), and 
the orbital extends in the direction perpendicular to the
Fe plane as $d_{x^2-y^2}$ in BaFe$_2$S$_3$ does.
So far, several studies have pointed out that
$d_{3z^2-r^2}$ in the 2D FeSC does not play a crucial role
in the magnetism and superconductivity.\cite{Capone1,Capone2, MisawaJPSJ, MisawaPRL}

\section{Two-orbital model and BZ unfolding}
\begin{figure}
\centering
\includegraphics[width=10cm]{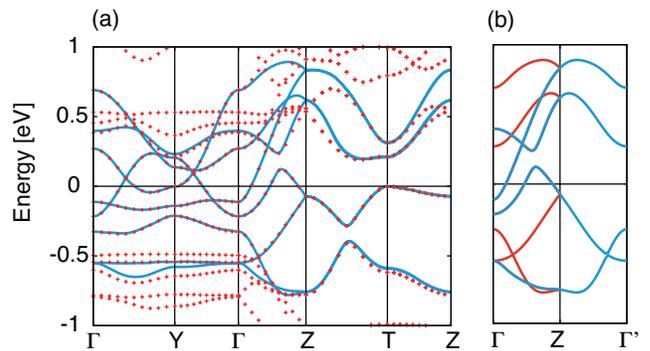}
\caption{(color online): (a) Band dispersion of the effective two-orbital (eight-band) 
model for $P$=0 (blue solid curves).
The original {\it ab initio} band structure is shown by dotted curves.
(b) Band dispersion in the extended (unfolded) BZ. 
}
\label{2orbital}
\end{figure}

We now further try to simplify the five-orbital (twenty-band) model. We have found that the energy dispersion 
around $E_F$ can be represented by a two-orbital (eight-band) model
in which the Wannier functions are mainly composed of two orbitals, i.e. $d_{x^2-y^2}$ and $d_{xz}$ hybridized with $d_{xy}$. We hereafter call the Wannier functions $w_{x^2-y^2}$ and $w_{xz}$. 
To construct the Wannier functions, the inner (outer) window is set to be [-0.3 eV,0.2 eV] ([-1.2eV,1.5 eV])
with respect to $E_F$.\cite{Souza}  
In Fig.~\ref{2orbital}(a), 
we compare the original {\it ab initio} bands and the Wannier interpolated bands for ambient pressure.
We obtained a similar result for $P$=12.4 GPa (not shown).
We list the resulting hopping integrals for the two-orbital model
in the Supplemental Material.\cite{suppl}

Interestingly, we can further simplify the current model by
unfolding the BZ, as was done for the 1111 compound.\cite{Kuroki}
Indeed, if we introduce a local gauge transformation for one of the two orbitals to change its sign, we can expand the band dispersion
from $\Gamma$ to $Z$ (see Fig.\ref{2orbital}(b)).\cite{downfoldnote} Then the model is simplified to a
four-band model.

\begin{figure}
\centering
\includegraphics[width=7cm]{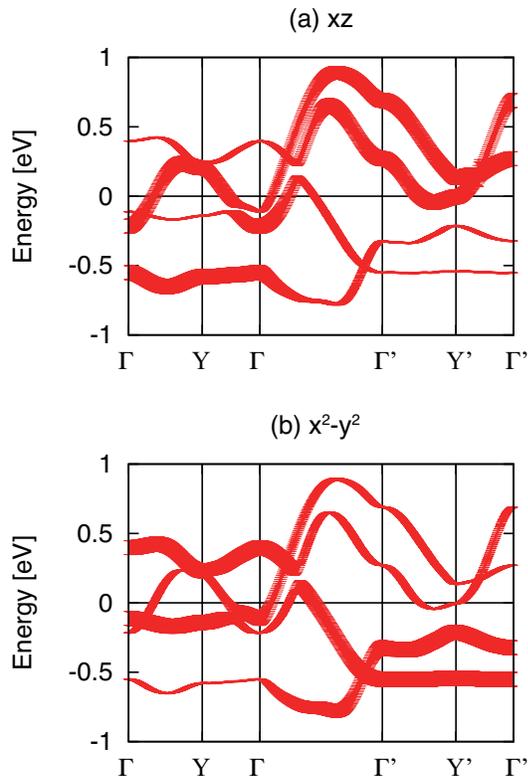}
\caption{(color online):
Band dispersion of the four-band model for $P$=0. The weight of the each orbitals
is orbital are represented by the width of the curve.
}
\label{fatband}
\end{figure}

In the 2D FeSC, both the hole and electron pockets
consist of more than two orbitals. Here let us look at the
orbital character of the Fermi surface of BaFe$_2$S$_3$.
In Fig.\ref{fatband}, we plot the band dispersion of the four-band model. 
The contribution of the each orbital is represented by the width of the curves.
We can see that $w_{xz}$ forms electron pockets around $\Gamma$
and $Y'$ (the Fermi pockets $\alpha$ and $\beta$ in the original BZ). On the other hand, 
the hole pocket sitting between $\Gamma$ and $\Gamma'$ (the $\gamma$ pocket in the
original BZ) 
consists of $w_{xz}$ and 
$w_{x^2-y^2}$. The far (near) side of the pocket from $\Gamma$ has a dominant  
$w_{x^2-y^2}$- ($w_{xz}$-) character.

\section{Lindhard function}
In the 2D FeSC, the Lindhard function $\chi^0({\bf q})$ has a peak around ($\pi$,0,0)
and (0,$\pi$,0) in the extended BZ\cite{Mazin}, 
which is compatible with the stripe-type antiferromagnetic order.
This fact has caused hot debates on which of the localized spin picture or
the itinerant picture is more appropriate for describing the magnetic properties of 2D
FeSC.\cite{DagottoNPhys,GreeneReview,Arita,Hansmann1,Hansmann2,BaFe2As2moment,Haule2}
Thus it is interesting to see whether the peak position of $\chi^0({\bf q})$ for BaFe$_2$S$_3$ agrees with
the Bragg peak observed in the experiment.

\begin{figure}
\centering
\includegraphics[width=7cm]{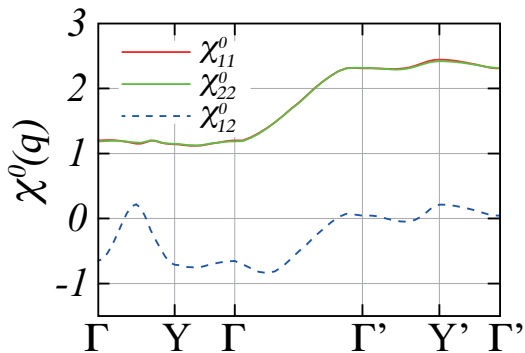}
\caption{(color online):
Wave number dependence of the Lindhard function for the five-orbital model at $P$=0.
}
\label{chi}
\end{figure}

In Fig.\ref{chi}, we plot $\chi^0_{ij}({\bf q})$ for the
five-orbital model in the extended Brillouin zone. Here $i$ and $j$(=1,2) specify the Fe atom in the unit cell. 
We see that the correlation along the ladder ($i=j$) is stronger than that along the rung $i\neq j$, and $\chi^0({\bf q})$
does not depend on $q_x$ and $q_y$ significantly.  As for the $q'_z$ dependence, 
$\chi^0({\bf q})$ has larger values in the plane of $q'_z=\pi$ than $q'_z=0$, due to the
nesting between the $\alpha$ and $\beta$ pockets in Fig.\ref{fermi}.
Note that the $\beta$ pocket moves to the plane of $q'_z=\pi$ in the
extended Brillouin zone. For $P$=12.4 GPa, although the $\beta$ pocket shrinks,
we confirmed that the contribution from the particle-hole scattering 
between $\alpha$ and $\beta$ is still dominant. 
This $q'_z$ dependence of $\chi^0({\bf q})$ is consistent with the experiment in that
the intra-orbital spin correlation is antiferromagnetic along the ladder.
As for the correlation along the rung, it is ferromagnetic, since 
ferromagnetically coupled spin configuration $S_1+S_2$ (where
$S_i$ is the spin at the $i$-th Fe atom in the unit cell) has stronger correlation than 
$S_1-S_2$ (note that $\chi^0_{12}>0$ at ${\bf q}=Y'$).
This is also consistent with the experiment.

On the other hand, for the inter-ladder correlation,  $\chi^0({\bf q})$ is peaked at $Y'$, while
the Bragg peak in experiment is observed to be located at 0.5${\bf b}_1$+0.5${\bf b}_2$
(${\bf b}_1$ and ${\bf b}_2$ are the reciprocal primitive vectors).\cite{Ohgushi}
This result suggests that the inter-ladder correlation can not be explained simply in terms of 
Fermi surface nesting.  

\section{Superconductivity}
Since the superconducting phase resides next to the antiferromagnetic phase, let us here
assume that the superconductivity is mediated by spin fluctuations. Then the pairing 
interaction is strong at $q'_z=\pi$, and favors a pairing gap function having different signs between 
the Fermi pockets $\alpha$ and $\beta$. 
The active band for superconductivity is $w_{xz}$ in this scenario, because
both $\alpha$ and $\beta$ are made from $w_{xz}$.
On the other hand, $w_{x^2-y^2}$ makes a pocket around $q'_z=\pm\pi/2$, but should be
passive for the superconductivity.
In the 2D FeSC, it has been proposed that $d_{xy}$ is important to
understand the material dependence of the superconducting transition 
temperature.\cite{Kuroki2,ScalapinoReview,MazinReview,PlattReview,ChubukovReview,DHLeeReview}
It is interesting to note that both $w_{xz}$ in BaFe$_2$S$_3$ and $d_{xy}$ in
the 2D FeSC extend in the direction of pnictogen/chalcogen sites, and
almost half-filled in the five-orbital model.

For 2D FeSC, the orbital fluctuation mechanism has been 
extensively studied.\cite{Kontani,Kontanivertex,Nomura} While it is an interesting future problem to
study whether orbital fluctuations mediate superconductivity in BaFe$_2$S$_3$, 
there is a notable difference between the quasi-1D and 2D systems.
In the case of the 2D FeSC,
$d_{yz}$ and $d_{xz}$ are essential for the orbital-fluctuation mechanism. While these orbitals
correspond to $d_{xy}$ and $d_{yz}$ in BaFe$_2$S$_3$, these are irrelevant in the
two-orbital model.

\section{Electron correlations}
Let us move on to the electron correlations in BaFe$_2$S$_3$. By means of the constrained random 
phase approximation (cRPA)\cite{cRPA}, we estimate the values of the Hubbard $U$ and 
the Hund's coupling $J$ in the five-orbital model. We used the density response code of Elk\cite{Elk,AntoncRPA}.
In the calculation of the charge susceptibility, we took 100 unoccupied bands and 4$\times$4$\times$4
{\bf k} and {\bf q} meshes. The double Fourier transform of the charge susceptibility was done with 
the cutoff of $|{\bf G}+{\bf q}|$=2,3,4,5 and 6 (1/a.u.) with ${\bf G}$ being the reciprocal vector, from which we estimated the values
extrapolated to the limit of $|{\bf G}+{\bf q}|\rightarrow \infty$.
In Table \ref{HubU}, we list the resulting values of $U$ and $J$ for $P=0$. 
If we compare these values (and the ratio $J/U$, which is important to measure the electron correlations in 2D FeSC\cite{Hundmetal})
with those of the 2D FeSC\cite{Miyake, Kazuma}, we see that they are as strong as those of LiFeAs. 
The pressure dependence of $U$ and $J$ is not so strong and they become smaller only by $\sim$ 6-7\%
under the pressures up to $P$=12.4 Pa. These results suggest that
BaFe$_2$S$_3$ is more strongly correlated than the 1111 compounds.\cite{Miyake,Haule} 
This observation is consistent with the fact that
the ordered moment of BaFe$_2$S$_3$ is about 1.3 $\mu_B$, which is larger than that of 
LaFeAsO\cite{magmomentLaFeAsO,Haule}.

\begin{table}
\caption{Hubbard $U$ and Hund's $J$ in the five-orbital model for $P$=0 GPa.}
\centering 
\begin{tabular}{c@{\,}r@{\,}r@{\,}r@{\,}r@{\,}r@{\ \ \ }r} \hline \hline
  & $U$ [eV] & $xy$ & $yz$& $3z^2$-$r^2$& $xz$& $x^2$-$y^2$ \\
  & & $\phantom{x^2-y^2}$ & $\phantom{x^2-y^2}$& $\phantom{x^2-y^2}$& $\phantom{x^2-y^2}$& $\phantom{x^2-y^2}$ \\[-3mm] \hline 
  &   $xy$      &\ 3.35& 2.42& 2.34& 2.33& 3.16\\   
  &   $yz$      &\ 2.42& 3.26& 2.89& 2.30& 2.54\\ 
  &   $3z^2$-$r^2$     &\ 2.34& 2.89& 3.79& 2.77& 2.44\\
  &   $xz$      &\ 2.33& 2.30& 2.77& 2.99& 2.43\\ 
  &$x^2$-$y^2$&\  3.16& 2.54&  2.44& 2.43& 3.86\\ \hline
\end{tabular}  
\begin{tabular}{c@{\,}r@{\,}r@{\,}r@{\,}r@{\,}r@{\ \ \ }r} \hline
  &$J$ [eV]       & $xy$ & $yz$& $3z^2$-$r^2$& $xz$& $x^2$-$y^2$ \\
  & & $\phantom{x^2-y^2}$ & $\phantom{x^2-y^2}$& $\phantom{x^2-y^2}$& $\phantom{x^2-y^2}$& $\phantom{x^2-y^2}$ \\[-3mm] \hline 
  &   $xy$      &\    -   & 0.50& 0.67& 0.48& 0.22\\   
  &   $yz$      &\ 0.50&   -   & 0.32& 0.46& 0.55\\ 
  &   $3z^2$-$r^2$     &\ 0.67& 0.32 &  - & 0.30& 0.71\\
  &   $xz$      &\ 0.48& 0.46 & 0.30 &  -  & 0.51\\ 
  &$x^2$-$y^2$&\  0.22& 0.55& 0.71& 0.51& -\\ \hline
\end{tabular} 
\label{HubU} 
\end{table}

\section{Conclusion}
We have studied the electronic structure of BaFe$_2$S$_3$ from first principles.
Detailed analysis of the full Fe orbital model revealed that the Fermi surfaces are represented only by the two 
Fe orbitals, i.e. $w_{xz}$ and $w_{x^2-y^2}$.
Provided that the superconductivity is mediated by spin fluctuations, $w_{xz}$,
which corresponds to $d_{xy}$ in the 2D FeSC, is essential for superconductivity.
In fact, they
have the following common features:
(i) they extend in the direction of 
pnictogen/chalcogen sites, (ii) they are nearly half-filled, (iii) they form 
disconnected Fermi pockets around $\Gamma$ and near the boundary of the BZ,
between which pairing interaction mediated by stripe-type antiferromagnetic
spin fluctuation works effectively.\cite{Disconn}
Concerning the electron correlations, it is stronger than those of LaFeAsO or LaFePO.


\begin{acknowledgments}
We thank K. Ohgushi, H. Takahashi,  J. Yamaura, and Y. Nambu for providing us 
their result\cite{Ohgushi,Ohgushi2} in prior to its publication and
Y. Nomura for useful discussions.
This work is financially supported by JSPS (Grants No.15H03696, RA; No.26800179, SS; 
No.15K17713, MS).
\end{acknowledgments}

\bibliographystyle{apsrev}
\bibliography{paper}

\begin{thebibliography}{66}
\bibitem[{dow()}]{downfoldnote}
\bibinfo{note}{The unfolded band is not exactly the same as the original band
  for general $k$ points. For another way to unfold the Brillouin zone, see
  e.g., Ref.\cite{Wei1,Wei3,Wei4,Wei5,Konbu}.}
\end{thebibliography}
\end{document}